\documentclass[lettersize,journal]{IEEEtran}

\usepackage{tabularx}

\ifCLASSINFOpdf
\else
\fi
\usepackage[numbers]{natbib}
\usepackage{epsfig}
\usepackage{epstopdf}
\usepackage[T1]{fontenc}
\usepackage{amssymb}
\usepackage{amsmath}
\usepackage{amsfonts}
\usepackage{verbatim}
\usepackage{graphicx}
\usepackage{hyperref}
\usepackage[usenames,dvipsnames]{xcolor}
\usepackage{lipsum}
\usepackage{booktabs}

\usepackage{multicol}
\usepackage{pgfplots}
\usepackage{lipsum}
\usepackage{colortbl}
\usepackage{caption}
\usepackage{algorithm}
\usepackage[noend]{algpseudocode}
\usepackage{enumitem}
\makeatletter
\def\algbackskip{\hskip-\ALG@thistlm}
\makeatother

\usepackage{multirow}

\usepackage{colortbl}
\usepackage{tabularx}
\usepackage{lipsum}  
\usepackage{pifont}
\usepackage{makecell}
\usepackage{float}
\usepackage{amsthm}
\usepackage{pgf}
\newcommand\subsubsubsection{\@startsection{paragraph}{4}{\z@}%
  {1.5ex \@plus 1ex \@minus .2ex}%
  {-1em}%
  {\normalfont\normalsize\bfseries}}
\makeatother

\usepackage{tikz,bm,color}

\usetikzlibrary{circuits.logic.US,circuits.logic.IEC,fit}

\begin{document}
%
\title{Energy-Efficient NTT Sampler for Kyber Benchmarked on FPGA}

\author{
  \IEEEauthorblockN{Paresh Baidya\IEEEauthorrefmark{1}\IEEEauthorrefmark{3}, Rourab Paul\IEEEauthorrefmark{2}, Vikas Srivastava\IEEEauthorrefmark{4}, Sumit Kumar Debnath\IEEEauthorrefmark{1}}\\
     \IEEEauthorblockA{\IEEEauthorrefmark{1}Department of Mathematics, National Institute of Technology, Jamshedpur, India}\\
     \IEEEauthorblockA{\IEEEauthorrefmark{2}Department of Computer Science and Engineering, Shiv Nadar University, Chennai, Tamil Nadu, India}\\
   \IEEEauthorblockA{\IEEEauthorrefmark{3}Department of Computer Science and Engineering, Siksha ‘O’ Anusandhan Deemed to be University, Bhubaneswar, India\\
        \IEEEauthorblockA{\IEEEauthorrefmark{4}Department of Mathematics, Indian Institute of Technology Madras, Chennai, India}\\      
          \IEEEauthorblockA{\IEEEauthorrefmark{3}pareshbaidya@soa.ac.in}
           \IEEEauthorblockA{\IEEEauthorrefmark{2}rourabpaul@snuchennai.edu.in}
           \IEEEauthorblockA{\IEEEauthorrefmark{4}vikas.math123@gmail.com}
           \IEEEauthorblockA{\IEEEauthorrefmark{1}sdebnath.math@nitjsr.ac.in}\\
   }
   
}
\maketitle
\vspace{-10pt}

\begin{abstract}
	Kyber is a lattice-based key encapsulation mechanism selected for standardization by the NIST Post-Quantum Cryptography (PQC) project. A critical component of Kyber's key generation process is the sampling of matrix elements from a uniform distribution over the ring~$\mathcal{R}_q$. This step is one of the most computationally intensive tasks in the scheme, significantly impacting performance in low-power embedded systems such as Internet of Things (IoT), wearable devices, wireless sensor networks (WSNs), smart cards, TPMs (Trusted Platform Modules), etc. Existing approaches to this sampling, notably conventional $\mathsf{SampleNTT}$ and {\sf Parse-SPDM3}, rely on rejection sampling. Both algorithms require a large number of random bytes, which needs at least three {\sf SHAKE-128} squeezing steps per polynomial. As a result, it causes significant amount of latency and energy. In this work, we propose a novel and efficient sampling algorithm, namely {\sf Modified SampleNTT}, which substantially reduces the average number of bits required from {\sf SHAKE-128} to generate elements in $\mathcal{R}_q$—achieving approximately a $33\%$ reduction compared to conventional {\sf SampleNTT}. {\sf Modified SampleNTT} achieves $99.16\%$ success in generating a complete polynomial using only two {\sf SHAKE-128} squeezes, outperforming both state-of-the-art methods, which never succeed in two squeezes of {\sf SHAKE-128}. Furthermore, our algorithm maintains the same average rejection rate as existing techniques and passes all standard statistical tests for randomness quality. FPGA implementation on Artix-7 demonstrates a $33.14\%$ reduction in energy, $33.32\%$ lower latency, and $0.28\%$ fewer slices compared to {\sf SampleNTT}. Our results confirm that {\sf Modified SampleNTT} is an efficient and practical alternative for uniform polynomial sampling in PQC schemes such as Kyber, especially for low-power security processors.
\end{abstract}

\vspace{-15pt}
\section{Introduction}
Public-key cryptographic systems play a fundamental role in ensuring secure communication over the internet by enabling encryption, digital signatures, and key exchange protocols. These protocols are primarily based on the hardness of mathematical problems such as integer factorization (RSA) and the discrete logarithm problem (Elliptic Curve Cryptography, ECC). However, the advent of quantum computing poses a significant threat to the security of these cryptosystems. When executed on a sufficiently powerful quantum computer, Shor’s algorithm can efficiently solve these problems in polynomial time, rendering traditional public-key cryptography insecure. Researchers have begun to study quantum-resistant public-key cryptographic algorithms to keep information secure from the upcoming quantum computer attack. This research area is known as Post-Quantum Cryptography (PQC). To address this emerging challenge, the U.S. National Institute of Standards and Technology (NIST) initiated the Post-Quantum Cryptography (PQC) standardization process in 2016 to develop quantum-resistant cryptographic algorithms. 
After four rounds of evaluation in 2022, NIST has decided to standardize CRYSTALS-Kyber as a key encapsulation mechanism (KEM) algorithm and CRYSTAL-Dilithium as a signature scheme. The theoretical foundation of the various lattice-based protocols is based on the computational hardness of Learning With Error (LWE). Its variants over Ring Learning With Error (RLWE) and \cite{rlwe} and Module Learning With Error (MLWE) \cite{mlwe} have been thoroughly investigated against both the classical and quantum adversaries.
A growing number of researchers are actively working on optimized implementations of PQC schemes on Central Processing Units (CPUs), Graphical Processing Units (GPUs), Application-Specific Integrated Circuits (ASICs) and Field-Programmable Gate Arrays (FPGAs), aiming to achieve optimal trade-offs in terms of resource usage, latency, and energy consumption across both software and hardware platforms. In particular, processor based software implementations offer flexibility and ease of deployment, whereas Application-Specific Integrated Circuits (ASICs) and reconfigurable hardware (e.g., FPGAs) provide high-performance and low-power solutions. FPGA-based implementations are becoming more popular compared to ASICs due to their cost-effectiveness and reconfigurable nature. In contrast, a pure hardware (FPGA) implementations can achieve significant performance by applying well-known optimization techniques such as register balancing, parallel processing, and efficient resource sharing. These designs are often reconfigurable and cost-effective, making them suitable for high-performance applications. 

CRYSTALS-Kyber is a lattice-based cryptosystem whose security relies on difficulty of solving the learning-with-errors (LWE) problem in module lattices. It is a quantum-resistant key encapsulation mechanism (KEM) that achieves IND-CCA2 (indistinguishability under adaptive chosen ciphertext attack) security. The fundamental algebraic operation in Kyber is of the form $\mathbf{A} \mathbf{s} + \mathbf{e}$, where $\mathbf{s}$ and $\mathbf{e}$ are $k$-dimensional polynomial vectors, and $\mathbf{A}$ is a $k \times k$ polynomial matrix with coefficients in the polynomial ring $R_q = \mathbb{Z}_{3329}[X]/(X^{256} + 1)$. Kyber supports three security levels - Kyber512, Kyber768, and Kyber1024—corresponding to NIST security levels 1, 3, and 5, respectively, with the module dimension $k$ set to 2, 3, and 4. 
To improve the practical applicability of Kyber, researchers have extensively focused on optimizing its computational efficiency and reducing latency across various algorithmic components. In the Kyber post-quantum cryptographic scheme, one of the most computationally intensive and resource-constrained operations are sampling, Number Theoretic Transform (NTT)-based polynomial multiplication, and cryptographic hashing. Wan et al.~\cite{Wan2022-vi} implemented a high-performance CRYSTALS-Kyber using an AI accelerator on a GPU (NVIDIA GeForce RTX 3080). They employ an NTT-box to perform the NTT/INTT operations efficiently, particularly when the polynomial dimension is relatively small. The implementation achieves a speedup of 6.47$\times$ compared to the state-of-the-art on the same GPU platform. In \cite{ji-HIkyber}, authors proposed three methods to improve the NTT performance: sliced layer merging (SLM), sliced depth first search (SDFS-NTT) and entire depth-first search (EDFS NTT). They uses kernel fusion-based memory optimization technique to achieve a speedup of 7.5\%, 28.5\%, and 41.6\% over the existing implementation. Notably, EDFS-NTT is introduced for the first time on a GPU platform (NVIDIA Titan V Volta GV100 with 5120 CUDA cores). Zou et al. \cite{zou-et-al} developed a RISC-V based processor: Seesaw, specifically designed to accelerate the Kyber schemes. This paper implemented the various components of the Kyber algorithm. Also, proposed a rejection sampling architecture where as, two coefficients are generated from the three uniform random bytes. Every three 4-bit random values pad with a 4-bit zeros to form a 16-bit word and is used to extract two matrix elements. In \cite{super-K-Kyber}, the paper proposed a dual-issue superscalar Kyber processor
(Super-K) based on RISC-V instruction set architecture (ISA). It is a parallel three stage pipeline architecture which supports conflict-free hash-based sampling and polynomial arithmetic operations. The authors \cite{super-K-Kyber} also design a reconfigurable polynomial arithmetic unit (PAU), which employs the fast modular reduction method and optimizes the compress/decompress process. This design reduced time overhead by 25\%–33\% and improved the parallelism and throughput of the overall processor. The work in \cite{Asic-1}, present a resource-efficient Kyber processor on ASIC (40nm LP CMOS) platform. The design incorporates a lightweight SHA-3 engine based on a half-fold Keccak core and reconfigurable modular arithmetic units (MAU) to compute the polynomial operations. The processor achieves a minimal power consumption of 273$\mu W$ and an energy efficiency of 0.72 $\mu J$ per operation in Kyber. Kim et al. \cite{kim-et-al} propose a configurable architecture for the Kyber accelerator, introducing a Memory-based Number Theoretic Transform (NTT) unit. The design also uses the Dadda tree algorithm for modular reduction, which improves processing speed, reduces hardware area, and increases data throughput.
The authors in \cite{ni2023towards}, present a lightweight BRAM-free FPGA implementation of the NTT/INTT unit in CRYSTALS-Kyber They propose an optimized modular multiplier based on K-RED~\cite{K-RED} and lookup-table techniques. The design outperforms existing works by 36–75\% in hardware efficiency and achieves a 3.4–4.4$\times$ improvement in point-wise multiplication performance. An instruction set coprocessor for Kyber is presented in \cite{bisheh} to design a high performance hardware architecture. This architecture also implements on ASIC platform which outperforms state-of-the-art implementations.
Article \cite{nguyen} implemented a Kyber using a non-memory-based iterative NTT for polynomial multiplication, which avoids the use of Block RAM on the Artix-7 FPGA. The authors in \cite{jati} implemented a light weight crypto processor for Kyber. Among all the aforementioned Kyber implementations, only \cite{nguyen} and \cite{jati} report the implementation cost of conventional \textsf{SampleNTT}.
\par A key computational process in Kyber is the generation of a structured public polynomial matrix in the Number Theoretic Transform (NTT) domain. These polynomials must be uniformly distributed over the ring $R_q$. The standard aforementioned implementations of Kyber utilize a rejection sampling technique in their \textsf{SampleNTT} algorithm, where random values are iteratively extracted from an input byte stream generated from {\sf SHAKE-128}, and out-of-range values are discarded until all polynomial coefficients are assigned valid values in $R_q$. To the best of our knowledge, all the aforementioned literature related to Kyber implementation has focused only on the NTT, polynomial multiplication, modules reduction, and the memory storage required to store polynomial coefficients and twiddle factors. The authors in \cite{spdm} proposed the first alternative to the rejection sampling algorithm for $Kyber$, adopting a simple partial discard method instead of the conventional rejection method used in \textsf{SampleNTT}. It reduces the required byte stream compared to results in \cite{kyber}. However, the article \cite{spdm} did not implement their sampler on a hardware,  thereby making it difficult to justify if this method will work in a practical implementation. Keep this in context, we propose a new \textsf{Modified SampleNTT} implemented on a hardware that not only drastically reduces energy consumption and latency but also conserves all the statistical properties of the conventional kyber \textsf{SampleNTT}.  The major contributions of our work are discussed below.

\subsection*{Our Contribution}

\begin{enumerate}
\item In this work, we propose a new sampling technique (namely \textsf{Modified SampleNTT}), that improves the efficiency of element generation in the polynomial ring $\mathcal{R}_q$ under the Kyber lattice-based cryptographic scheme. In particular, \textsf{Modified SampleNTT} offers hardware efficiency improvements over the conventional \textsf{SampleNTT} used in Kyber. It achieves a 33.14\% reduction in energy consumption, 33.32\% lower latency, and a 0.28\% decrease in slice utilization. All implementations were performed and verified using the Vivado 22.04 tool on the Artix-7 FPGA platform.

\item  We demonstrate that \textsf{Modified SampleNTT} reduces the average number of bits required to generate an element in $\mathcal{R}_q$ when the \textsf{XOF} is instantiated with \textsf{SHAKE-128}. Across all Kyber security levels (Kyber512, Kyber768, and Kyber1024), our method consistently consumes only $\sim2523.8$ bits, compared to $\sim3785$ bits in conventional \textsf{SampleNTT} and $\sim3470$ bits in \textsf{Parse-SPDM3}. The proposed algorithm successfully generates a full polynomial in $\mathcal{R}_q$ using exactly {two squeezing steps} of \textsf{SHAKE-128} in {99.16\%} of the cases. In contrast, existing approaches such as conventional \textsf{SampleNTT} and \textsf{Parse-SPDM3} fail to achieve this, consistently requiring three or more invocations of {\sf SHAKE-128}.

\item Furthermore, \textsf{Modified SampleNTT} maintains the same average rejection sampling percentage as the existing designs, \textsf{SampleNTT} and \textsf{Parse-SPDM3}, across all Kyber security levels (as shown in Table~\ref{tab:rejection-sampling-percentage}). The rejection rate remains consistently within the narrow range of {18.84\% to 18.85\%}.
	
\item In addition, \textsf{Modified SampleNTT} passes all standard statistical tests for randomness, including the Frequency, Entropy, Kolmogorov–Smirnov (KS), Wald–Wolfowitz, and Serial tests (see Table~\ref{tab:statistical-tests}). These results confirm that  \textsf{Modified SampleNTT} retains the quality of uniform sampling and exhibits randomness characteristics on par with state-of-the-art algorithms like conventional \textsf{SampleNTT} and \textsf{Parse-SPDM3}.

\end{enumerate}
The organization of the article is as follows: 
Section \ref{sec:pre} presents the preliminaries and problem statement. The proposed \textsf{Modified SampleNTT} and its analysis are detailed in Sections \ref{sec:mod:sampleNTT} and \ref{sec:ana}, respectively. The hardware architecture and corresponding results are discussed in Sections \ref{sec:hw} and \ref{sec:hw:dis}. Finally, the conclusions are presented in Section \ref{sec:con}.
\section{Preliminaries}
\label{sec:pre}

\begin{algorithm}
\caption{Kyber.\textbf{CPAPKE.KeyGen}(): Key Generation \cite{kyber}}
\begin{algorithmic}[1]
\State {\bf Output}: Secret key $sk \in \mathbb{B}^{12 \cdot k \cdot n / 8}$
\State {\bf Output}: Public key $pk \in \mathbb{B}^{12 \cdot k \cdot n / 8 + 32}$
\State $d \leftarrow \mathbb{B}^{32}$
\State $(\rho, \sigma) := G(d)$
\State $N := 0$
\For{$i$ from $0$ to $k - 1$}
    \For{$j$ from $0$ to $k - 1$}
        \State $\hat{A}[i][j] := \textsf{Parse}(\textsf{XOF}(\rho, j, i))$
    \EndFor
\EndFor
\State \textellipsis
\end{algorithmic}
\end{algorithm}

\begin{algorithm}
\caption{\textsf{SampleNTT}: $\mathbb{B}^* \to R_q$ \cite{kyber}}
\label{algo:samplentt}
\begin{algorithmic}[1]
\State \textbf{Input:} Byte stream $B = \beta_0, \beta_1, \beta_2, \dots \in \mathbb{B}^*$
\State \textbf{Output:} NTT-representation $\hat{a} \in R_q$ of $a \in R_q$
\State $i := 0$
\State $j := 0$
\While{$j < n$}
    \State $d_1 := \beta_i + 256 \cdot (\beta_{i+1} \mod^+ 16)$
    \State $d_2 := \lfloor \beta_{i+1} / 16 \rfloor + 16 \cdot \beta_{i+2}$
    \If{$d_1 < q$}
        \State $\hat{a}_j := d_1$
        \State $j := j + 1$
    \EndIf
    \If{$d_2 < q$ \textbf{and} $j < n$}
        \State $\hat{a}_j := d_2$
        \State $j := j + 1$
    \EndIf
    \State $i := i + 3$
\EndWhile
\State \Return $\hat{a}_0 + \hat{a}_1 X + \dots + \hat{a}_{n-1}X^{n-1}$
\end{algorithmic}
\end{algorithm}

\subsection{Public Key Generation in Kyber: $\mathsf{ SampleNTT}$}

The key generation algorithm of {\sf Kyber} produces a secret key $sk \in \mathcal{B}^{12 \cdot k \cdot n / 8}$ and a public key $pk \in \mathcal{B}^{12 \cdot k \cdot n / 8 + 32}$. To derive a public key, the algorithm first constructs a matrix $\hat{{A}} \in R_q^{k \times k}$ within the \textit{NTT} (Number Theoretic Transform) domain. This matrix is generated by invoking an algorithm named \textsf{SampleNTT} $k^2$ times. The \textsf{SampleNTT} algorithm processes a byte stream $B = b_0, b_1, b_2, \dots \in \mathcal{B}^*$ and produces the \textit{NTT} representation $\hat{a} = \hat{a}_0 + \hat{a}_1 X + \dots + \hat{a}_{n-1} X^{n-1} \in R_q$ of $a \in R_q$. The input byte stream for Algorithm \textsf{SampleNTT} is generated using \textsf{XOF}. The function \textsf{XOF} is recommended to be instantiated with {\sf SHAKE-128}.

To sample an element uniformly from $R_q$, Algorithm \textsf{SampleNTT} extracts twelve-bit chunks sequentially from the input byte stream and assigns a coefficient to the element only if the extracted chunk falls within the required range. If the chunk is out of range, it discards the chunk and extracts another twelve-bit segment, following a simple rejection sampling method. This process repeats until all coefficients of the element in $R_q$ are determined. Consequently, the number of bytes (or bits) required by Algorithm \textsf{SampleNTT} to generate $\hat{a} \in R_q$ is not fixed and depends on the input byte stream.

\section{$\mathsf{Modified SampleNTT}$}
\label{sec:mod:sampleNTT}
In this section, we present the design of \textsf{Modified SampleNTT} algorithm (refer Algorithm \ref{algo:modified-sample-ntt}). The proposed algorithm is an optimized polynomial sampling method that is more efficient compared to the \textsf{SampleNTT} algorithm \ref{algo:samplentt} used in \textsf{Kyber}. The \textsf{SampleNTT} algorithm employs a rejection sampling technique, takes a byte stream as input to generate the coefficients of the polynomial in $R_q$.
In this paper, {\sf Modified SampleNTT} reduces the number of required byte streams compared to \textsf{SampleNTT}. It ensures a consistent sample rejection rate and efficient use of randomness. {\sf Modified SampleNTT} takes an input byte stream $B$, and converts it into a polynomial $ \hat{\alpha} \in R_q$. It extracts \textit{twelve-bit chunks} at a time and maps them into polynomial coefficients. Two twelve-bit values, $ d_1 $ and $d_2 $, are extracted from two consecutive bytes in each iteration. In line [7-8], The first value, $d_1$, is computed by combining the $i^{th}$ byte and the next $(i+1)^{th}$ byte using bitwise OR operation. Then, apply a 12-bit mask to take the relevant bits to make a number in the range [0, 4095]. Similarly, $d_2$  is derived but with the byte order reversed to ensure the randomness in the output. In line 9 and line 12, the calculated values $ d_1 $ and $d_2 $ are compared against the modulus q = 3329 respectively;if the value is less than $q$, it is accepted as a valid coefficient; otherwise it is rejected.

\begin{algorithm}
\caption{\textsf{Parse-SPDM3}: $\mathbb{B}^* \to R_q$ \cite{spdm}}
\label{algo:spdm}
\begin{algorithmic}[1]
\State \textbf{Input:} Byte stream $B = \beta_0, \beta_1, \beta_2, \dots \in \mathbb{B}^*$
\State \textbf{Output:} NTT-representation $\hat{a} \in R_q$ of $a \in R_q$
\State $i := 0$
\State $j := 0$
\While{$j < n$}
    \State $d_1 := 16\cdot \beta_i + \lfloor \beta_{i+1}/ 16 \rfloor$
    
    \If{$d_1 < 3584$}
        \State $i := i + 1$
    \EndIf
    \State $d_2 := (256 \cdot  \beta_{i}~ mod^{+}~ 2^{12}) + \beta_{i+1}$
    \If{$d_1 < q $}
        \State $\hat{a}_j := d_1$
        \State $j := j + 1$
    \EndIf
    \If{$d_2 < q$ \textbf{and} $j < n$}
        \State $\hat{a}_j := d_2$
        \State $j := j + 1$
    \EndIf
    \If{$d_2 < 3584$}
        \State $i := i + 2$
    \EndIf
    \State $i := i + 1$
\EndWhile
\State \Return $\hat{a}_0 + \hat{a}_1 X + \dots + \hat{a}_{n-1}X^{n-1}$
\end{algorithmic}
\end{algorithm}

\begin{algorithm}
\caption{\textsf{Modified SampleNTT}: $\mathbb{B}^* \to R_q$}
\label{algo:modified-sample-ntt}
\begin{algorithmic}[1]
\State \textbf{Input:} Byte stream $B = \beta_0, \beta_1, \beta_2, \dots \in \mathbb{B}^*$
\State \textbf{Output:} NTT-representation $\hat{a} \in R_q$ of $a \in R_q$
\State $i := 0$
\State $j := 0$
\State $\textsf{mask} := 4095$
\While{$j < n$}
    \State $d_1 := ((\beta_i \;|\; (256 \cdot \beta_{i+1})) \;\&\; \textsf{mask}$\label{line:d1}
    \State $d_2 := ((\beta_{i+1} \;|\; 256 \cdot \beta_i)\; \&\; \textsf{mask}$  \label{line:d2}
    \If{$d_1 < q$}
        \State $\hat{a}_j := d_1$
        \State $j := j + 1$
    \EndIf
    \If{$d_2 < q$ \textbf{and} $j < n$}
        \State $\hat{a}_j := d_2$
        \State $j := j + 1$
    \EndIf
    \State $i := i + 2$
\EndWhile
\State \Return $\hat{a}_0 + \hat{a}_1 X + \dots + \hat{a}_{n-1}X^{n-1}$
\end{algorithmic}
\end{algorithm}


\section{Analysis}
\label{sec:ana}
We first present the comparative analysis of the average number of bits required to generate an element in $\mathcal{R}_q$ when {\sf XOF} is instantiated with {\sf SHAKE-128} across three Kyber security levels: Kyber512, Kyber768, and Kyber1024 (refer Table \ref{tab:avg-no-of-bits-required}). This metric quantifies the amount of randomness extracted from the {\sf XOF}, which in turn determines the entropy and computational effort involved in generating uniform elements in the ring. We compare the conventional {\sf SampleNTT}, and {\sf Parse-SPDM3} with {\sf Modified SampleNTT}. For each security level, {\sf Modified SampleNTT} consistently achieves a significant reduction in the number of bits required—approximately $2523.8$ bits—compared to conventional {\sf SampleNTT}, which requires over $3785$ bits, and {\sf Parse-SPDM3}, which uses around $3470$ bits. The experiments were carried out over $1$ million iterations. These findings confirm that {\sf Modified SampleNTT} provides a more resource-efficient approach to element generation in $\mathcal{R}_q$.  Reducing the average number of bits needed per element in $\mathcal{R}_q$ leads to less expansion of {\sf XOF} expansion, lower energy consumption and faster execution. Thus, {\sf Modified SampleNTT} method is suitable for practical deployment in resource-constrained cryptographic environments.

\begin{table*}[]
	\centering
	\caption{{Average numbers of bits required to generate an element in $\mathcal{R}_q$ when {\sf XOF} is instantiated with {\sf SHAKE-128}}}
	\scalebox{1.5}{
		\begin{tabular}{@{}lccc@{}}
			\toprule
			\textbf{Cipher} & \textbf{{\sf SampleNTT}} & {{\sf Parse-SPDM3}} & {\sf Modified SampleNTT} \\
			\midrule
			\textbf{Kyber512}  & 3785.8446  & 3470.3500  & 2523.8184 \\
			\textbf{Kyber768}  & 3785.7993  & 3470.3177  & 2523.8201 \\
			\textbf{Kyber1024} & 3785.8269  & 3470.3223 & 2523.8084 \\
			\bottomrule
	\end{tabular}}
    \label{tab:avg-no-of-bits-required}
\end{table*}
Table~\ref{tab:success-percetage-two-calls-hash} presents an analysis of the {success percentage for generating an entire polynomial in} $\mathcal{R}_q$ {using exactly two squeezing steps of the {\sf SHAKE-128}}. {\sf Modified SampleNTT} achieves a {\sf 99.16\%} success rate in generating a full polynomial using only two {\sf SHAKE-128} invocations, in contrast to the {\it 0\% } success rate of both conventional {\sf SampleNTT} and {\sf Parse-SPDM3}. In addition, both {\sf SampleNTT} and {\sf Parse-SPDM3} require at least {three or more invocations} of {\sf SHAKE-128} to produce enough randomness for polynomial generation. Thus, they incur additional computational cost, latency, and power consumption. In contrast, {\sf Modified SampleNTT} optimally utilized the {\sf XOF} output, and reduce the number of hash function calls required to sample a polynomial in $\mathcal{R}_q$. 

\begin{table*}[]
	\centering
	\caption{Success percentage that an element in $\mathcal{R}_q$ is generated using exactly two {\sf SHAKE-128} squeezing steps}
	\scalebox{1.5}{
		\begin{tabular}{@{}ccc@{}}
			\toprule
		 \textbf{{\sf SampleNTT}} & {{\sf Parse-SPDM3}} & {\sf Modified SampleNTT} \\
			\midrule
			 0\%  & 0\%  & 99.16\% \\
			
			\bottomrule
	\end{tabular}}
        \label{tab:success-percetage-two-calls-hash}

\end{table*}
\begin{table*}[]
	\centering
	\caption{{Average rejection percentage while generating an element in $\mathcal{R}_q$ when {\sf XOF} is instantiated with {\sf SHAKE-128}}}
	\scalebox{1.5}{
		\begin{tabular}{@{}lccc@{}}
			\toprule
			\textbf{Cipher} & \textbf{{\sf SampleNTT}} & {{\sf Parse-SPDM3}} & {\sf Modified SampleNTT} \\
			\midrule
			\textbf{Kyber512}  & 18.84\%  & 18.85\%  & 18.84\% \\
			\textbf{Kyber768}  & 18.85\%  & 18.84\%  & 18.85\% \\
			\textbf{Kyber1024} & 18.81\% & 18.82\% & 18.85\% \\
			\bottomrule
	\end{tabular}}
    \label{tab:rejection-sampling-percentage}
\end{table*}
Table~\ref{tab:rejection-sampling-percentage} presents the \textit{average rejection percentage} while generating an element in $\mathcal{R}_q$ when the {\sf XOF} is instantiated with {\sf SHAKE-128}. Rejection sampling plays a key role in ensuring uniformity of the generated elements. {\sf Modified SampleNTT} maintains rejection percentages that are statistically equivalent to those of conventional {\sf SampleNTT} and {\sf Parse-SPDM3} across all Kyber security levels. We note that our improvements in bit consumption (Table~\ref{tab:avg-no-of-bits-required}) and reducing the number of hash calls of {\sf SHAKE-128} (Table~\ref{tab:success-percetage-two-calls-hash}) do not compromise sampling correctness and efficiency.

\subsection{Statistical Analysis }
In order to  evaluate the statistical quality and randomness of the sampled output sequences generated by {\sf Modified SampleNTT}, a comprehensive benchmarking of well-established randomness tests has been conducted. These tests are designed to uncover different types of non-random patterns or statistical anomalies. We also conducted these randomness test on existing state-of-the-art sampling algorithm such as conventional {\sf SampleNTT} and {\sf Parse-SPDM3} for the completeness. Each subsection introduces the theoretical motivation and mathematical formulation of the respective test, followed by a detailed analysis of the results observed across the three sampling algorithms. 
 \subsubsection{Frequency Test}
 
 The Frequency Test is used to analyze the uniformity of a sequence of discrete random variables. The  objective of this test is to determine whether all possible symbols in the output sequence of a sampling algorithm appear with approximately equal frequency, as expected in a truly uniform random process. Given a sequence $S = \{s_1, s_2, \dots, s_n\}$ of $n$ values sampled from a discrete set $\{0, 1, \dots, k-1\}$, let $O_i$ denote the observed frequency of symbol $i$ and let the expected frequency under the uniform distribution be $E = n/k$ for all $i = 0, 1, \dots, k-1$. In the following, the observed frequencies are compared with the expected ones using the chi-square ($\chi^2$) statistic, defined as
 \[
 \chi^2 = \sum_{i=0}^{k-1} \frac{(O_i - E)^2}{E}.
 \]
Under the null hypothesis $H_0$ that the data is uniformly distributed, the $\chi^2$ statistic follows a chi-square distribution with $(k - 1)$ degrees of freedom. To determine whether the observed sequence deviates significantly from uniformity, we computed the $p$-value associated with the observed $\chi^2$ statistic. If this $p$-value is smaller than a pre-determined significance level $\alpha = 0.05$, the null hypothesis is rejected, indicating that the sequence does not exhibit uniform randomness. It is also important to consider the standard deviation of the frequencies, which gives an auxiliary metric of spread around the mean frequency $E$. The standard deviation $\sigma$ of the frequencies is defined as:
 \[
 \sigma = \sqrt{ \frac{1}{k} \sum_{i=0}^{k-1} (O_i - E)^2 },
 \]
 A low standard deviation means values occur with nearly equal frequencies, indicating uniformity, while a high value suggests potential non-randomness. All three algorithms were tested using $25,600,000$ samples drawn over a value space of size $k = 3329$. For an ideal uniform distribution, the expected frequency per symbol is approximately $7689.997$, with a theoretical standard deviation around $87.66$. The {\sf Modified SampleNTT} method yielded a mean of $7690.00$ and a standard deviation of $87.66$, producing a chi-square statistic of $3326.6658$ with a p-value of $0.503265$, thereby passing the test at the $0.05$ significance level. Similarly, the standard {\sf SampleNTT} method resulted in a chi-square statistic of $3342.4143$ and a p-value of $0.426774$, with a standard deviation of $87.87$, also passing the test. The {\sf Parse-SPDM3} method exhibited slightly higher variability, with a standard deviation of $88.85$, a chi-square value of $3417.3402$, and a p-value of $0.137072$, but still comfortably passed the threshold for statistical uniformity. These results confirm that all three algorithms exhibit near-uniform output distributions and demonstrate no statistically significant deviation from ideal randomness as assessed by the frequency test and the chi-square goodness-of-fit method.

\subsubsection{Serial Test}

The Serial Test (also known as the two-dimensional frequency test) is a statistical tool that is used for the evaluation of the uniformity and independence of adjacent elements in a sequence. Let a sequence $S = \{x_1, x_2, \dots, x_n\}$ consist of $n$ samples where each $x_i$ takes values from a finite set $\mathcal{A} = \{0, 1, \dots, k-1\}$. In the following, we construct ordered pairs $(x_1, x_2), (x_2, x_3), \dots, (x_{n-1}, x_n)$ resulting in $n - 1$ adjacent pairs. Under the hypothesis that $S$ is generated by a truly uniform and independent process, each possible pair $(a, b) \in \mathcal{A} \times \mathcal{A}$ should appear with approximately equal probability, i.e., $\frac{1}{k^2}$. Since, there are $n-1$ total pairs, the expected count for each such pair is given by
\[
E = \frac{n - 1}{k^2},
\]
Let $O_{i,j}$ denote the observed frequency of the pair $(i,j)$ in the sequence. We employ the chi-square statistic to measure the deviation of observed pair frequencies from their expected values
\begin{align}
\chi^2 = \sum_{i=0}^{k-1} \sum_{j=0}^{k-1} \frac{(O_{i,j} - E)^2}{E}\label{test:st-1}.	
\end{align}
Equation \ref{test:st-1} approximately follows a chi-square distribution with $k^2 - 1$ degrees of freedom under the null hypothesis that the sequence is uniformly distributed and consecutive elements are independent. The corresponding $p$-value is calculated using the CDF of the chi-square distribution. A small $p$-value (typically $p < 0.01$ or $p < 0.05$) indicates that the observed distribution of pairs is significantly different from the expected uniform distribution. For all three sampling algorithms {\sf Modified SampleNTT}, {\sf SampleNTT}, and {\sf Parse-SPDM3}—the chi-square statistic was approximately $11081986$, with $11082240$ degrees of freedom. This yielded a consistent $p$-value of $\approx 0.521457$ in each case. Since the $p$-value exceeds the common significance level of 0.05, the results indicate that all three algorithms produce sequences that exhibit randomness in terms of serial correlation.

\subsubsection{Runs Test}

The Runs Test evaluate the randomness of a binary sequence by analyzing the occurrence and distribution of uninterrupted subsequences (called ``runs'') of similar elements. In other words, the test checks whether the number and lengths of such runs are consistent with what would be expected in a truly random sequence, where each bit is independently and uniformly distributed. Let us consider a binary sequence $S = \{x_1, x_2, \dots, x_n\}$ of length $n$, where each $x_i \in \{0,1\}$. A \emph{run} is defined as a maximal contiguous subsequence of identical bits. For example, in the sequence $S = 00110011$, there are four runs: two of $0$s and two of $1$s, with various lengths. We assume that the bits are independent and identically distributed with equal probability of 0 or 1. Let $n_0$ and $n_1$ denote the total number of $0$s and $1$s, respectively, in the sequence. The total number of runs, denoted $R$, can be determined by iterating through the sequence and incrementing the run count each time a change in bit value is observed. Let $\mu_R$, and $\sigma_R^2$ denotes respectively, the expected number of runs under the hypothesis of randomness, and the corresponding variance. Using these parameters, the test statistic is computed as a standard normal variable:
\[
Z = \frac{R - \mu_R}{\sigma_R}.
\]
Under the null hypothesis that the sequence is random, the value of $Z$ follows a standard normal distribution. A $p$-value is then computed from $Z$, and the null hypothesis is rejected if this $p$-value falls below a pre-determined threshold, such as 0.05 or 0.01. By mapping integer sequences to binary (e.g., thresholding), one can apply the test in our case.

For the {\sf Modified SampleNTT} implementation, the observed number of runs was $499470$ against an expected value of $500000.9997$, yielding a $Z$-score of $-1.0620$ and a $p$-value of $0.4882$. The conventional {\sf SampleNTT} implementation showed $499777$ runs against expected $500000.8330$, with a $Z$-score of $-0.4477$ and a $p$-value of $0.3544$. Lastly, the {\sf Parse-SPDM3} variant produced $500621$ runs against the expected $500000.9804$, resulting in a $Z$-score of $1.2400$ and a $p$-value of $0.2150$. In all three cases, the $p$-values were well above the conventional significance threshold of 0.05, leading to the conclusion that the null hypothesis of randomness could not be rejected. Thus, the output sequences from all three protocols are consistent with what would be expected from a random source in terms of run behavior.

\subsubsection{KS Test}
The Kolmogorov–Smirnov (KS) Test is a non-parametric statistical method used to assess the goodness-of-fit between an empirical distribution function (EDF) derived from a given data sample and a reference CDF. In other words, KS test can be used to test whether a sample of data conforms to a specified probability distribution. The EDF of the observed data is compared to the CDF of the theoretical distribution. The test statistic, denoted $D_n$, is defined as the supremum of the absolute differences between these two functions over the entire range of the data: $D_n = \sup_x |F_n(x) - F(x)|$, where $F_n(x)$ is the EDF and $F(x)$ is the CDF of the reference distribution. In the case of a two-sample K-S test, the test compares two empirical distributions $F_n(x)$ and $G_m(x)$ from independent samples, using the test statistic $D_{n,m} = \sup_x |F_n(x) - G_m(x)|$. To determine statistical significance, the computed test statistic is used to compute a p-value. A smaller KS statistic value suggests a closer match to the reference distribution. We took the reference distribution to be uniform. For all three sampling algorithms {\sf Modified SampleNTT}, {\sf SampleNTT}, and {\sf Parse-SPDM3}—the KS statistics were $0.00208$, $0.00082$, and $0.00106$ respectively, with corresponding $p$-values of $0.7777$, $0.5073$, and $0.2041$. Since all $p$-values are greater than the common significance threshold of 0.05, the null hypothesis that the sample follows the reference distribution cannot be rejected in any case. These results confirm that the outputs of all three protocols conform closely to the uniform distribution and can be considered statistically random under the KS test.

\subsubsection{Entropy Test}

The Entropy Test quantifies the level of uncertainty or unpredictability in a sequence of discrete random variables. The motivation behind this test is based on the idea that a truly random sequence should exhibit maximum entropy. 
Mathematically, for a discrete random variable $X$ with a finite set of outcomes $\{x_1, x_2, \dots, x_k\}$ and corresponding empirical probabilities $p_i = \Pr(X = x_i)$ for $i = 1, 2, \dots, k$, the Shannon entropy $H(X)$ is defined as:
\begin{align}
H(X) = -\sum_{i=1}^{k} p_i \log_2 p_i\label{test:entropy}.
\end{align}

This measure reaches its maximum value when all outcomes are equally likely, i.e., when $p_i = 1/k$ for all $i$, resulting in $H_{\text{max}} = \log_2 k$. We consider a sequence $S = \{s_1, s_2, \dots, s_n\}$ consisting of $n$ symbols drawn from an alphabet of size $k$. The test involves counting the occurrences of each symbol $x_i$ in $S$ to estimate their empirical probabilities $p_i = \frac{f_i}{n}$, where $f_i$ is the frequency count of $x_i$. The entropy of the sequence is then computed using the formula above. This value is then compared to the theoretical maximum entropy $\log_2 k$ expected from a perfectly uniform random distribution. A small deviation from the maximum indicates good randomness, while a significant drop suggests potential bias or pattern structure in the data. We utilized the entropy test to assess the unpredictability in the sequence outputted by {\sf SampleNTT}, {\sf Modified SampleNTT}, and {\sf Parse-SPDM3}. The entropy \( H \) of a discrete probability distribution over a set of size \( k = 3329 \) is maximized when all possible outcomes are equally likely, in which case the expected entropy value is \( H_{\text{expected}} = \log_2(3329) \approx 11.7009 \). For each of the three sampling schemes, the observed entropy was computed using Equation \ref{test:entropy}. All three methods yielded nearly identical empirical entropy values of \( H = 11.7008 \), which are extremely close to the theoretical maximum.

\begin{table*}[h!]
	\centering
	\caption{\textbf{Statistical Test Results for Randomness }}
	\label{tab:statistical-tests}
	\scalebox{1.3}{
	\begin{tabular}{@{}lccc@{}}
		\toprule
		 & {\sf SampleNTT} & {\sf Parse-SPDM3} & {\sf Modified SampleNTT} \\
		\midrule
		{ Frequency Test} & {\color{black}\ding{51}} & {\color{black}\ding{51}} & {\color{black}\ding{51}} \\
		{ Entropy Test} & {\color{black}\ding{51}} & {\color{black}\ding{51}}  & {\color{black}\ding{51}} \\
		{ Kolmogorov-Smirnov (KS) Test}  & {\color{black}\ding{51}}  & {\color{black}\ding{51}}  & {\color{black}\ding{51}} \\
		{ Wald-Wolfowitz Test} & {\color{black}\ding{51}} & {\color{black}\ding{51}} & {\color{black}\ding{51}} \\
		{ Serial Test} & {\color{black}\ding{51}} & {\color{black}\ding{51}}  & {\color{black}\ding{51}} \\
		\bottomrule
	\end{tabular}}
\end{table*}
\section{Hardware Architecture}
\label{sec:hw}
This section discusses the hardware architectures of conventional {\sf SampleNTT} and {\sf Modified SampleNTT}.

\subsection{Conventional {\sf SampleNTT}}
\begin{figure}[!htb]
\centering
\includegraphics[width=0.5\textwidth]{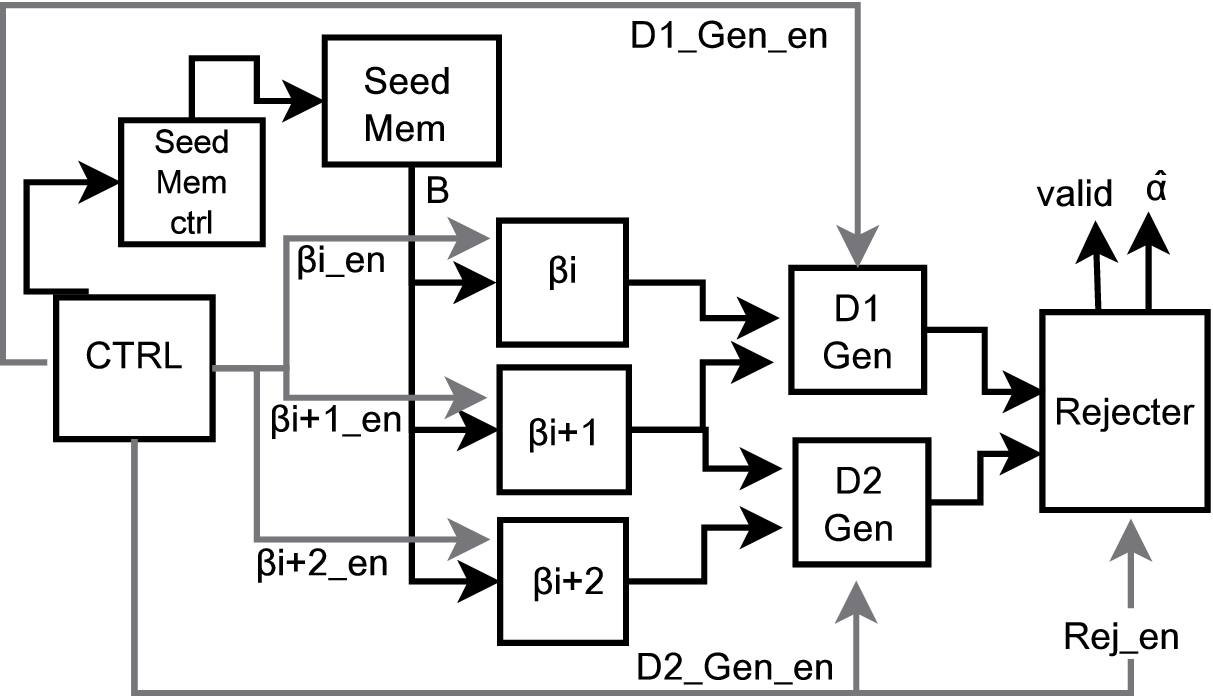}
\vspace{-5pt}
\caption{Hardware Architecture of Conventional SampleNTT used in Kyber }
\vspace{-5pt}
\label{fig:sampleNTT}
\end{figure}

As shown in Fig. \ref{fig:sampleNTT}, the conventional {\sf SampleNTT} used in Kyber  has 9 sub components.
\subsubsection{Seed Memory ($SeedMem$)}
Seed Memory ($SeedMem$) is a First In First Out (FIFO) memory which stores $504$ bytes generated by {\sf SHAKE-128} algorithm used in Kyber  variants. This block has $7$ ports : $clk\_rd$, $clk\_wr$, $rd\_en$, $wr\_en$, $rst$, $din$ and $dout$. The $SeedMem\_ctrl$ generates the control signals : $rd\_en$, $wr\_en$, $rst$  of $SeedMem$ to read the output $B=[\beta_0, \beta_1, \beta_2, ...]$ from $SeedMem$. The output of $SeedMem$ is buffered in $\beta_{i}$ $Block$, $\beta_{i+1}$ $Block$ and $\beta_{i+2}$ $Block$ blocks.
\subsubsection{Seed Memory Controller $SeedMem\_ctrl$}
This $SeedMem\_ctrl$ module generates control signals : $rd\_en$, $wr\_en$, $rst$ for the $SeedMem$. Once the conventional {\sf SampleNTT} is enabled, $SeedMem\_ctrl$ reads bytes from $SeedMem$ on each rising clock edge and sends them to the $\beta_{i}$, $\beta_{i+1}$, and $\beta_{i+2}$ blocks.
\subsubsection{Controller ($CTRL$)}
The Controller ($CTRL$) block generates enable ($en$) and reset ($rst$) signals for $SeedMem\_ctrl$, $\beta_i$, $D1\_Gen$, $D2\_Gen$, $Rejecter$ $block$, $\beta_{i+1}$ $block$ and $\beta_{i+2}$ blocks. 
\subsubsection{$\beta_{i}$ Block}
The $\beta_{i}$ block latches the $0_{th}$, $3_{rd}$, $6_{th}$, ... bytes from $SeedMem$ when $\beta_{i}\_en$ from $CTRL$ is high.

\subsubsection{$\beta_{i+1}$ Block}
The $\beta_{i+1}$ block latches the $1_{st}$, $4_{th}$, $7_{th}$, ... bytes from $SeedMem$ when $\beta_{i+1}\_en$ from $CTRL$ is high.
\subsubsection{$\beta_{i+2}$ Block}
The $\beta_{i+2}$ block latches the $2_{nd}$, $5_{th}$, $8_{th}$, ... bytes from $SeedMem$ when $\beta_{i+2}\_en$ from $CTRL$ is high.
\subsubsection{D1 Generator ($D1\_Gen$)}
When $CTRL$ sets $D1\_Gen\_en$ high, the D1 Generator ($D1\_Gen$) reads $\beta_{i}$ from the $\beta_{i}~Block$ and $\beta_{i+1}$ from the $\beta_{i+1}~Block$, then executes line 6 of Algorithm \ref{algo:samplentt}.
\subsubsection{D2 Generator ($D2\_Gen$)}
When $CTRL$ sets $D2\_Gen\_en$ high, the D2 Generator ($D2\_Gen$) reads $\beta_{i+1}$ from the $\beta_{i+1}~Block$ and $\beta_{i+2}$ from the $\beta_{i+2}~Block$, then executes line 7 of Algorithm \ref{algo:samplentt}.
\subsubsection{Rejecter Block ($Rej\_Block$)}
When $CTRL$ sets $Rej\_en$ high, the Rejecter Block ($Rej\_Block$) checks whether $D1$ and $D2$ are less than $q$ or not (line 8 and line 11 \ref{algo:samplentt}). If this condition holds true, $D1$ and $D2$ are accepted; otherwise, they are rejected.
\begin{figure}[!htb]
\centering
\includegraphics[width=0.5\textwidth]{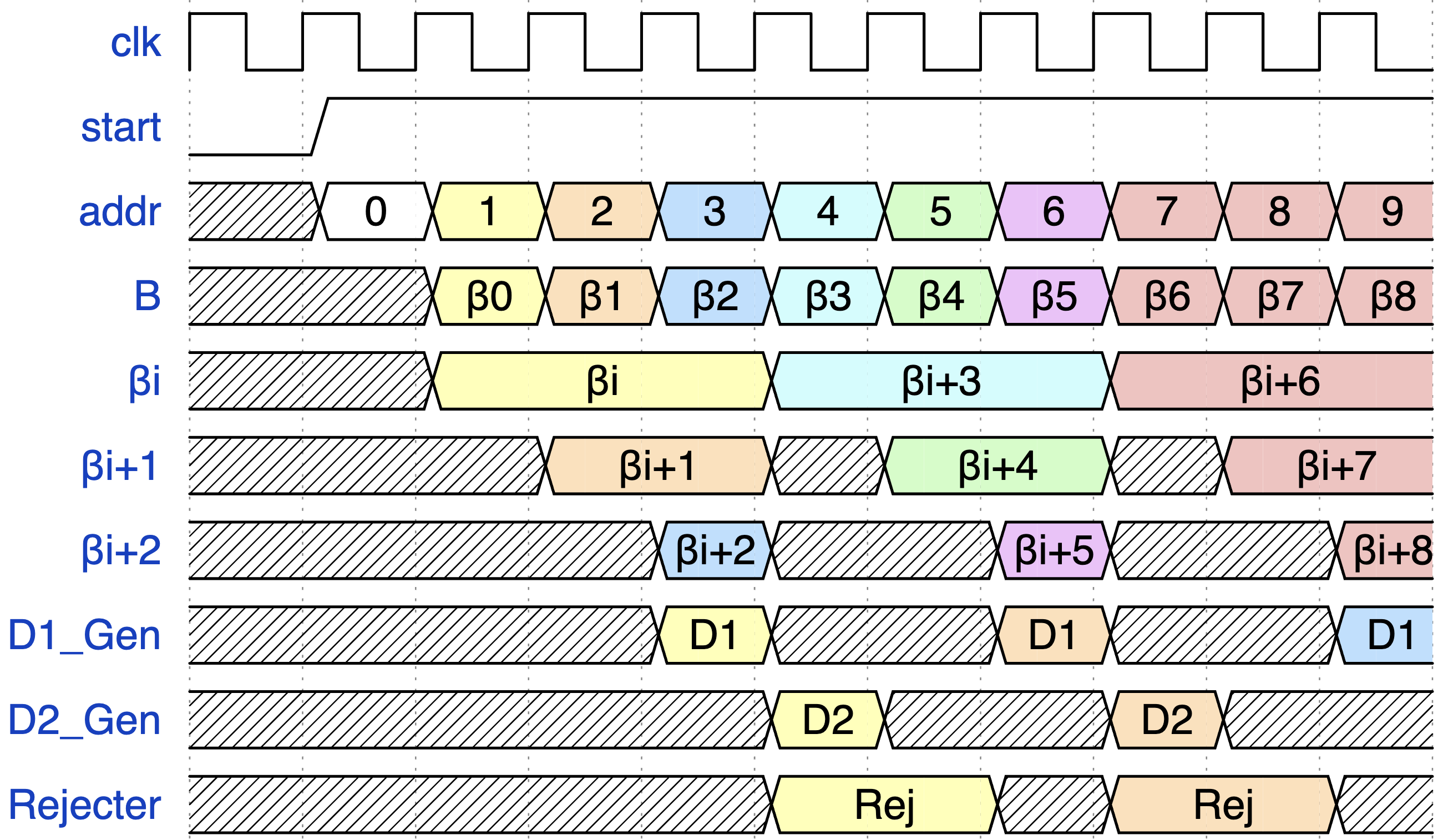}
\vspace{-5pt}
\caption{Timing Diagram of Conventional {\sf SampleNTT} used in Kyber }
\vspace{-5pt}
\label{fig:timing_sampleNTT}
\end{figure}
\par Fig. \ref{fig:timing_sampleNTT} shows the timing diagram of conventional {\sf SampleNTT}. The $\beta_0$, $\beta_1$, $\beta_2$, .. are read in every clock cycle from $SeedMem$. 
The values $\beta_0$, $\beta_3$, $\beta_6$, \ldots are latched by the $\beta_i$ $Block$ every 3 clock cycles.
Similarly, $\beta_1$, $\beta_4$, $\beta_7$, \ldots are latched every 3 clock cycles by the $\beta_{i+1}$-Block, and $\beta_2$, $\beta_5$, $\beta_8$, \ldots are latched every 3 clock cycles by the $\beta_{i+2}$-Block. The $\beta{i}$, $\beta{i+1}$ and $\beta{i+2}$ are latched for 3 clock cycles, 2 clock cycles and 1 clock cycles respectively. Once $\beta_{i}$ and $\beta_{i+1}$ are latched by the $\beta_{i}$ Block and $\beta_{i+1}$ Block, respectively, the $D1\_Gen$ is ready to compute $D1$. Similarly, once $\beta_{i+1}$ and $\beta_{i+2}$ are latched by the $\beta_{i+1}$ Block and $\beta_{i+2}$ Block, respectively, the $D2_Gen$ is ready to compute $D2$. The $Rejector$ $Block$ starts once $D1$ is computed and compares $D1$ with $q$. After $D1$ is ready, $D2$ is computed in the next clock cycle. The $Rejector$ $Block$ then continues by comparing $D2$ with $q$. As a result, the $Rejector$ $Block$ is activated in every 3-clock-cycle period. For the first 2 clock cycles, the $Rejecter$ $Block$ remains active, and for the remaining 1 clock cycle, it is inactive.
\subsection{Proposed {\sf Modified SampleNTT}}
\begin{figure}[!htb]
\centering
\includegraphics[width=0.5\textwidth]{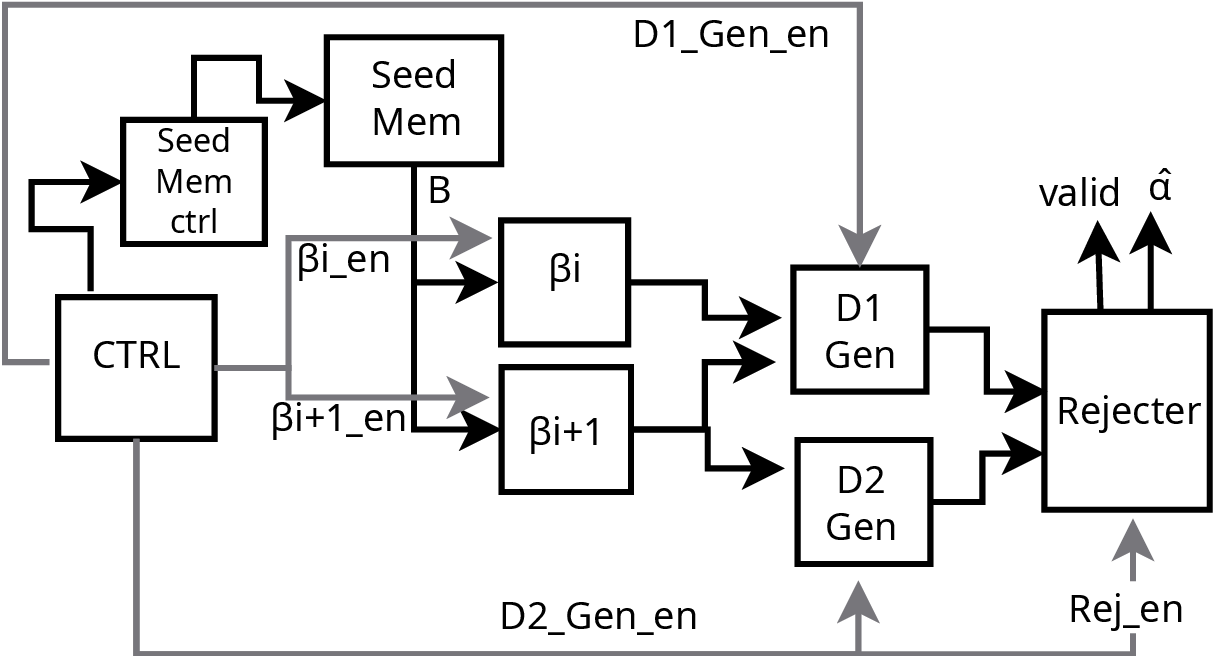}
\vspace{-5pt}
\caption{Hardware Architecture of Modified Sample NTT}
\vspace{-5pt}
\label{fig:arch_sampleNTT}
\end{figure}

The proposed modified {\sf SampleNTT} stated in algorithm \ref{algo:modified-sample-ntt} does not required $\beta_{i+2}$ $Block$. As shown in Fig. \ref{fig:arch_sampleNTT}, this {\sf Modified SampleNTT} has three primary changes.
\subsubsection{$\beta_{i+2}$ $Block$} The modified {\sf SampleNTT} does not require $\beta_{i+2}$ $Block$. The $D1$ and $D2$ can be computed directly from $\beta_{i}$ and $\beta_{i+1}$.
\subsubsection{CTRL} As $\beta_{i+2}$ $Block$ is absent, the control signals of $\beta_{i+2}$ $Block$ from $CTRL$ are not required. Therefore, it reduces the implementation cost of $CTRL$.
\subsubsection{D1 \& D2 Generator ($D1\_Gen$ \& $D2\_Gen$)}
As shown in lines \ref{line:d1} and \ref{line:d2} of Algorithm \ref{algo:modified-sample-ntt}, apart from the $OR$ operation, both lines involve a multiplication by $256$ and an $AND$ operation with the $mask$ value $4095$.To make the hardware more efficient, we replace these two operations with two lightweight operations.
\begin{itemize}
\item The multiplication by $256$ is replaced by the left-shifting $\beta_{i+1}$ (line~\ref{line:d1}) and $\beta_{i}$ (line~\ref{line:d2}) by 8 bits.

\item The mask operation with $4095$ is replaced by truncating all bits beyond the 12th bit (since $2^{12} = 4096$).
\end{itemize}
\subsubsection{Seed Memory}
The modified {\sf SampleNTT} requires only $\sim 336$ bytes to store in $SeedMem$ for Kyber standard, while the conventional {\sf SampleNTT} requires $\sim 504$ bytes for storage in $SeedMem$. 

\begin{figure}[!htb]
\centering
\includegraphics[width=0.5\textwidth]{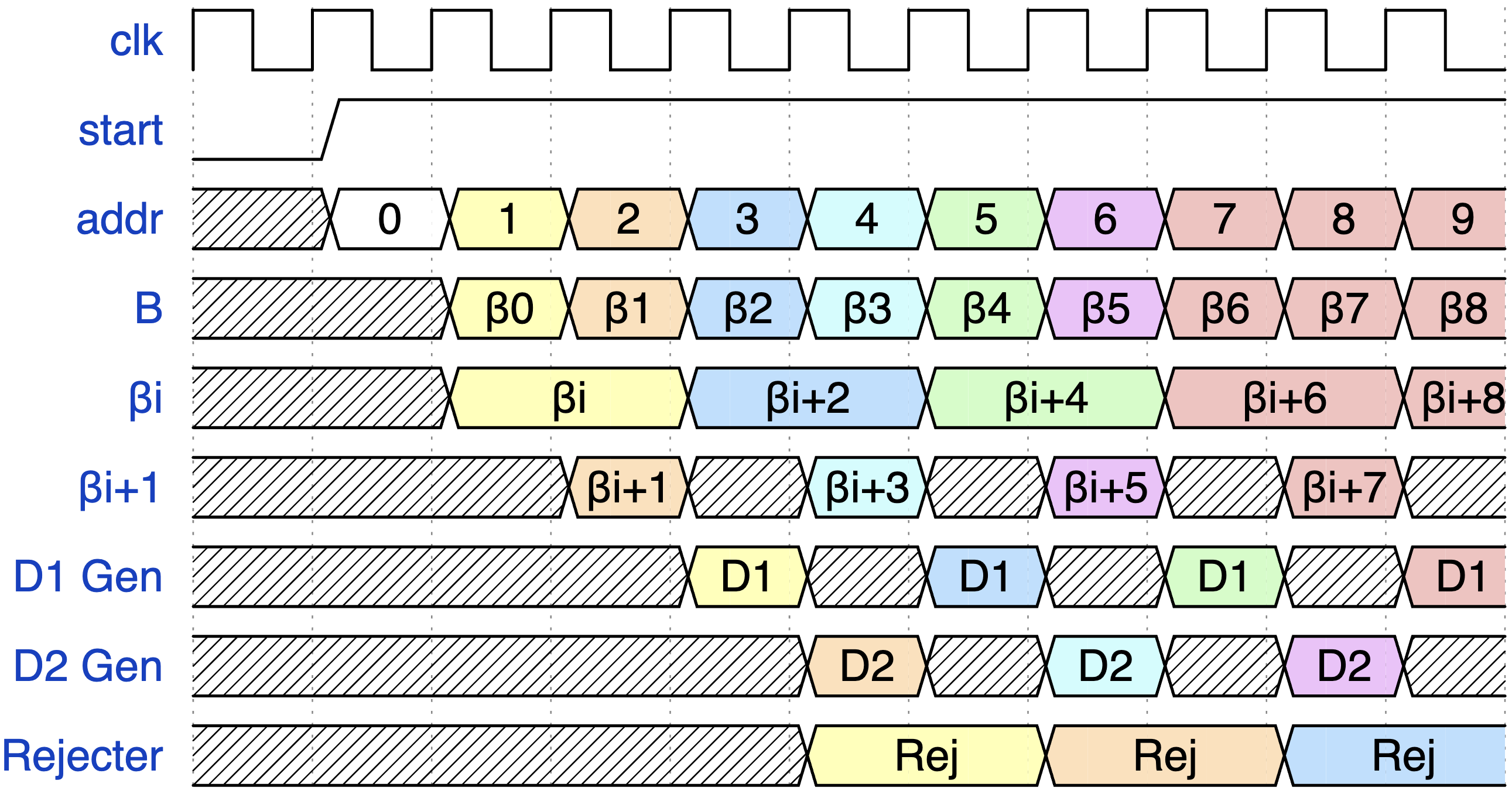}
\vspace{-5pt}
\caption{Timing Diagram of {\sf Modified SampleNTT}s}
\vspace{-5pt}
\label{fig:mod_timing_sampleNTT}
\end{figure}
\begin{table*}[!htbp]
 	\centering
    \resizebox{\textwidth}{!}{
	\begin{tabular}{|>{\raggedright\arraybackslash}p{3.6cm}|>{\centering\arraybackslash}p{1.0cm}|>{\centering\arraybackslash}p{1.0cm}|>{\centering\arraybackslash}p{1.0cm}|>{\centering\arraybackslash}p{1.15cm}|>{\centering\arraybackslash}p{1.15cm}|>{\centering\arraybackslash}p{1.4cm}|>{\centering\arraybackslash}p{1.15cm}|}
	\hline
    \textbf{Design Names} &\textbf{\# Slices} & \textbf{\# LUTs} & \textbf{\# FFs} & \textbf{\# DSPs} & \textbf{Energy (nJ)} & \textbf{Clock Period (ns)} & \textbf{\# Clock Cycles}\\ \hline\hline
    \textbf{Sample NTT \cite{nguyen}} &\textbf{62} &\textbf{116} & \textbf{141} & \textbf{0} &\textbf{NA*} & \textbf{NA*} & \textbf{NA*}  \\ 
         \hline
                \hline 
    \textbf{Sample NTT \cite{jati}} &\textbf{78} &\textbf{246} & \textbf{133} & \textbf{0} &\textbf{NA*} & \textbf{3.7} & \textbf{4623}  \\ 
         \hline
                \hline 
	\textbf{Sample NTT $^\star$} &\textbf{44} &\textbf{113} & \textbf{96} & \textbf{0} &\textbf{$\sim$470} & \textbf{10} & \textbf{$\sim$474}  \\ 
	            - SeedMem\_ctrl      &9   &15  & 20  &0  &     &  &     \\ 
	            - $\beta_{i}$ Block &1   &0   & 8   &0     &     &   &    \\ 
	            - $\beta_{i+1}$ Block &3   &0   & 8   &0     &     &   &    \\
	            - $\beta_{i+2}$ Block &1   &0   & 8   &0     &     &  &     \\
	            - d1\_gen        &4   &6   & 12  &0     &     &  &     \\
	            - d2\_gen        &7  &13  & 12  &0    &     &  &     \\
	            - rejecter       &9   &3   & 15  &0     &     & &      \\
	            - seed\_mem      &22  &72  &10  &0    &     &  &     \\
	            - CTRL     &6   &4 & 6 &0  &  & &    \\
               {\sf SHAKE-128 $^\circledast$}   &\textbf{4115}& \textbf{37577} & \textbf{47898} &0     & \textbf{$\sim$3114}  & \textbf{3.264}  &\textbf{3324} \\ \hline
                \hline 

\textbf{SPDM3 $^\star$} &\textbf{40} &\textbf{106} & \textbf{79} & \textbf{0} &\textbf{$\sim$320} & \textbf{10} & \textbf{$\sim$316}  \\ 
	            - SeedMem\_ctrl      &6   &9  & 19  &0  &     &  &     \\ 
	            - $\beta_{i}$ Block &2   &0   & 8   &0     &     &   &    \\ 
	            - $\beta_{i+1}$ Block &3   &0   & 8   &0     &     &   &  \\
	            - d1\_gen        &9   &14   & 12  &0     &     &  &     \\
	            - d2\_gen        &4  &4  & 12  &0    &     &  &     \\
	            - rejecter       &10   &4   & 15  &0     &     & &      \\
	            - seed\_mem      &22  &72  &10  &0    &     &  &     \\
	            - CTRL     &3   &3 & 6 &0  &  & &    \\
                {\sf SHAKE-128 $^\circledast$}   &\textbf{4115}& \textbf{37577} & \textbf{47898} &\textbf{0}     & \textbf{$\sim$3114}  & \textbf{3.264}  &\textbf{3324} \\ 
                \hline \hline
\textbf{Our Modified Sample NTT $^\star$}  &\textbf{32} & \textbf{106} & \textbf{79} &\textbf{0} &\textbf{$\sim$320} & \textbf{10} &\textbf{$\sim$316} \\ 
		        - SeedMem\_ctrl      &5  & 9 & 20 &0     &     & & \\ 
	            - $\beta_{i}$ Block &2  &0 & 8 &0     &     &  &     \\ 
             	- $\beta_{i+1}$ Block &3  &0 & 8 &0     &     &  &     \\
             	- seed\_mem      &23 &72 & 10 &0     &     &  &     \\
             	- d1\_gen      & 7 &14 & 15 &0     &     &  &     \\
                    - d2\_gen      & 4 &4 & 15 &0     &     &  &     \\
                     - rejecter       &9   &4   & 15  &0     &     & &      \\
             	- CTRL        &2  &3 & 3 & 0    &  &   &       \\
                    {\sf SHAKE-128 $^\circledast$}   &\textbf{4115}& \textbf{37577} & \textbf{47898} &\textbf{0 }    & \textbf{$\sim$2076 } & \textbf{3.264 } &\textbf{2216} \\ 
                
\hline\hline
                \multicolumn{8}{|c|}{\textbf{The proposed {\sf Modified SampleNTT$^\star$} is successfully tested with Kyber-512 on the Artix-7 FPGA (xc7a100tcsg324-3).}} \\ \hline
\end{tabular}}
\vspace{2pt}
\caption*{Note: NA* = Data Not Available; $^\star$ = Designed by us; $^\circledast$ = AMD-Xilinx Vitis Security Library \cite{vitis:sec}.}

 		\caption{Implementation Costs of Sample NTTs}
 	\label{tab:impl}
 \end{table*}

\begin{figure}[!htb]
\centering
\includegraphics[width=0.5\textwidth]{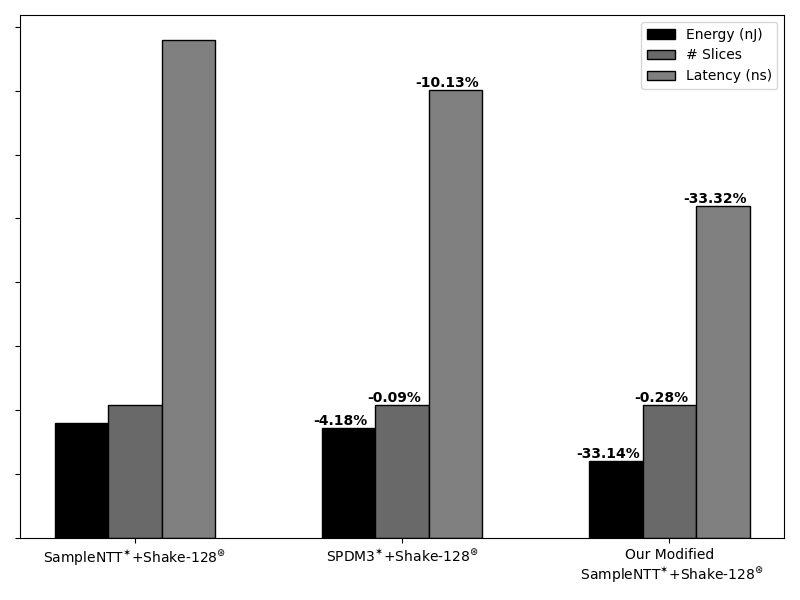}
\vspace{-5pt}
\caption{Reduction of Implementation Cost of Our {\sf Modified SampleNTT$^\star$} + ${\sf SHAKE-128}^\circledast$ and {\sf SPDM3$^\star$} + ${\sf SHAKE-128}^\circledast$  as Compared to {\sf SampleNTT$^\star$} + ${\sf SHAKE-128}^\circledast$}
\vspace{-5pt}
\label{fig:allcost}
\end{figure}

 \begin{figure}[!htb]
\centering
\includegraphics[width=0.5\textwidth]{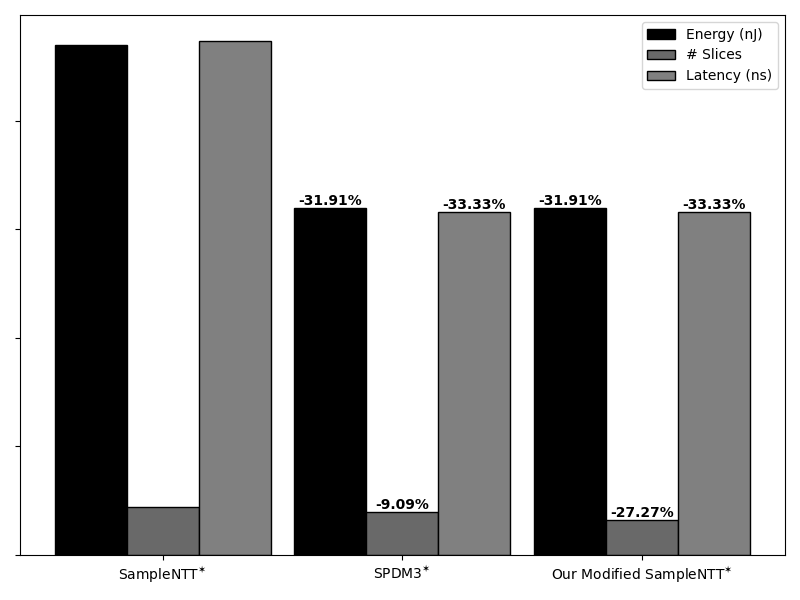}
\vspace{-5pt}
\caption{Reduction of Implementation Cost of Our {\sf Modified SampleNTT$^\star$} and {\sf SPDM3$^\star$} as Compared to {\sf SampleNTT$^\star$}}
\vspace{-5pt}
\label{fig:cost}
\end{figure}

\section{Hardware Results \& Discussions }
\label{sec:hw:dis}
The conventional {\sf SampleNTT} used in Kyber, {\sf Parse-SPDM3}  and proposed {\sf Modified SampleNTT} are implemented on $Artix$-$7$ $(xc7a100tcsg324$-$3)$ FPGA with the $Vivado$ $22.02$ tool and the VHDL language. This section discusses the impacts of the adopted changes on the implementation cost of our {\sf Modified SampleNTT}. The detailed implementation costs of our conventional {\sf SampleNTT}$^\star$, the {\sf SampleNTT} from \cite{nguyen}, the {\sf SampleNTT} from \cite{jati}, {\sf Parse-SPDM3}$^\star$ from \cite{spdm}, and the proposed {\sf Modified SampleNTT}$^\star$ are presented in Table~\ref{tab:impl}. It is to be noted that the $^\star$ symbols indicate the hardware designs are implemented by us. To the best of our knowledge, articles \cite{nguyen} and \cite{jati} are the only works in the literature that reported the implementation cost of the {\sf SampleNTT}.
\subsubsection{Impact on Energy and Time}
Our proposed modified {\sf SampleNTT$^\star$} can generate the required number of $d1$s and $d2$s using $336$ input bytes generated from the ${\sf SHAKE-128}^\circledast$, whereas the conventional {\sf SampleNTT$^\star$} and {\sf Parse-SPDM3$^\star$} require $504$ bytes from the ${\sf SHAKE-128}^\circledast$. These extra bytes generated in ${\sf SHAKE-128}^\circledast$ for both the conventional {\sf SampleNTT$^\star$} and the {\sf Parse-SPDM3$^\star$} require extra clock cycles, which causes a significant amount of latency and energy. As a result, as shown in Fig. \ref{fig:allcost}, our {Modified \sf SampleNTT$^\star$} alone efficiently reduces energy consumption by $31.91\%$ and latency by $33.33\%$, compared to the conventional {\sf SampleNTT$^\star$} used in Kyber. Our {Modified \sf SampleNTT$^\star$}+${\sf SHAKE-128}^\circledast$ efficiently reduces energy consumption by $33.14\%$ and latency by $33.32\%$, compared to the {\sf SampleNTT$^\star$}+${\sf SHAKE-128}^\circledast$ used in Kyber.
On the other hand, {\sf Parse-SPDM3$^\star$} is able to reduce energy consumption by $4.18\%$ and latency by $10.13\%$, compared to the {\sf SampleNTT$^\star$} used in Kyber. It is to be noted that the main {\sf SampleNTT$^\star$} and ${\sf SHAKE-128}^\circledast$ can run in parallel in all the 3 designs. However, the latencies of {\sf Parse-SPDM3$^\star$}, the conventional {\sf SampleNTT$^\star$} and our {\sf Modified SampleNTT$^\star$} are calculated as the sum of the latencies of {\sf SampleNTT$^\star$}/{\sf Parse-SPDM3$^\star$} and ${\sf SHAKE-128}^\circledast$.
 The detailed timing diagram of conventional {\sf SampleNTT} used in Kyber  and our {\sf Modified SampleNTT}s are shown in Fig. \ref{fig:timing_sampleNTT} and Fig. \ref{fig:mod_timing_sampleNTT} respectively. As the {\sf SampleNTT} used in Kyber  requires 3 bytes from $SeedMem$ to generate one set of $d_1$, $d_2$, there is an empty clock cycle after generating each set. However, our modified {\sf SampleNTT} requires only 2 bytes from $SeedMem$ to generate one set of $d_1$, $d_2$, and thus does not incur any empty clock cycles. As a result, considering the rejection rate to generate 256 sets of $d_1$, $d_2$, our {\sf SampleNTT} requires $\sim 316$ clock cycles, whereas the conventional {\sf SampleNTT} requires $\sim 474$ clock cycles. In our modified design, using fewer bytes from $Shake$-128 and eliminating the empty clock cycle in {\sf SampleNTT} significantly reduces both energy consumption and latency.
\subsubsection{Impact on Area}
Our modified {\sf SampleNTT$^\star$} eliminates the $\beta_{i+2}$ blocks. Therefore, our modified {\sf SampleNTT$^\star$} can produce the required number of $d_1$ and $d_2$ using only $336$ bytes generated from ${\sf SHAKE-128}^\circledast$, instead of the $504$ bytes required by the conventional {\sf SampleNTT$^\star$}. As a result, it reduces the FIFO depth of the $SeedMem$ used to store the bytes from ${\sf SHAKE-128}^\circledast$. Additionally, the $CTRL$ block of our {\sf SampleNTT$^\star$} becomes lightweight as it no longer needs to generate control signals for the $\beta_{i+2}$ block. As a result, the slice consumption of our modified {\sf SampleNTT$^\star$} is reduced by $27.27\%$ compared to the conventional {\sf SampleNTT$^\star$} used in Kyber, whereas {\sf Parse-SPDM3$^\star$} achieves only a $9.09\%$ reduction in slice overhead relative to conventional{\sf SampleNTT$^\star$}.\\
\textbf{Availability of Codes}\\ The RTL and the statistical test code for this work are uploaded to GitHub \footnote{https://github.com/rourabpaul1986/SampleNTT}.
\section{Conclusion}
\label{sec:con}
{\sf SampleNTT} and $Shake-128$ are among the most fundamental and critical components of next-generation security processors incorporating Kyber. However, {\sf SampleNTT} and $Shake-128$ cause significant energy and latency overhead, which makes them unsuitable for low-power field embedded systems. {\sf SampleNTT} requires an adequate number of random bytes from $Shake-128$ in Kyber. Generating more random bytes using $Shake-128$ leads to increased latency and energy consumption. To make Kyber suitable for low-power, resource-constrained devices, the proposed {\sf Modified SampleNTT} adopts two measures without affecting its statistical properties: (i) It reduces the required number of random bytes form $Shake-128$ and (ii) The proposed algorithm for {\sf Modified  SampleNTT} avoids extra buffering of random byte and develops light wight Controller. As a result, our {\sf Modified SampleNTT}+$Shake-128$ reduces energy consumption by $33.14\%$, latency by $33.32\%$ and slice utilization by $0.28\%$, compared to the {\sf SampleNTT}+$Shake-128$ used in Kyber. Meanwhile, our {\sf Modified SampleNTT} alone reduces energy consumption by $31.91\%$, latency by $33.32\%$ and slice utilization by $27.27\%$ compared to the {\sf SampleNTT} used in Kyber.  As part of future work, we aim to further reduce the rejection sampling percentage while preserving uniformity and correctness. Additionally, an in-depth study of the rejection sampling behaviour in other lattice-based schemes such as Dilithium will be performed to explore the broader applicability of our optimization techniques.

\bibliographystyle{unsrt}  
\bibliography{IEEEexample}

\end{document}